\documentclass[12pt]{article}

\usepackage[a4paper,margin=2.5cm]{geometry}
\usepackage[utf8]{inputenc}
\usepackage[T1]{fontenc}
\usepackage[brazilian]{babel}
\usepackage{amsmath,amssymb,amsthm}
\usepackage{graphicx}
\usepackage{booktabs}
\usepackage{multirow}
\usepackage{natbib}
\usepackage{ragged2e}
\usepackage{threeparttable}
\usepackage{float}
\usepackage{placeins}
\usepackage{setspace}
\usepackage{newtxtext,newtxmath}
\usepackage{microtype}
\usepackage{caption}
\usepackage{xurl}
\usepackage{hyperref}
\hypersetup{
  colorlinks=true,
  linkcolor=blue,
  citecolor=blue,
  urlcolor=black,
  pdftitle={Jornada de trabalho, produtividade e informalidade: uma avaliacao quantitativa para o Brasil},
  pdfauthor={}
}
\graphicspath{{./}}
\setcitestyle{authoryear,round,aysep={}}
\renewcommand{\abstractname}{Resumo}

\newcommand{\tabnotes}[1]{\par\vspace{2pt}\noindent\begin{minipage}{\linewidth}\footnotesize #1\end{minipage}}
\newcommand{\Areq}{A_{\mathrm{req}}}

\newif\ifanonymous
\anonymoustrue

\title{\textbf{Jornada de trabalho, produtividade e informalidade: uma avalia\c{c}\~ao quantitativa para o Brasil}}

\author{
    Victor Rangel\\
    {\small Insper}
    \and
    Fernando Barros Jr\\
    {\small FEA-RP USP}
}
\date{}

\let\paperbibliography\thebibliography
\renewcommand{\thebibliography}[1]{\paperbibliography{#1}\widowpenalty=10000\clubpenalty=10000}
\begin{document}
\maketitle
\vspace{-1.0cm}

\begin{abstract}
\noindent Este artigo quantifica o ganho de produtividade total dos fatores (PTF) necessário para neutralizar, no curto prazo, a perda de produto associada a tetos alternativos para a jornada semanal no Brasil. Combinamos momentos da PNAD Contínua de 2024T4 com um modelo estático de equilíbrio parcial no qual trabalho formal e informal são substitutos imperfeitos, a eficiência horária depende da jornada e as cunhas de formalização governam a composição do emprego. Para representar as reformas de 44 para 40 e 36 horas, limitamos previamente as horas habituais formais a 44h, como hipótese de mensuração da base. A alternativa de 40 horas requer $1{,}53$--$1{,}57\%$ de PTF. A passagem para 36 horas requer entre $6{,}52\%$ e $8{,}38\%$, conforme a hipótese sobre eficiência abaixo de 40 horas; no mesmo contrafactual, a informalidade aumenta entre $2{,}28$ e $2{,}92$ pontos percentuais. Decomposições mostram que a contração física de horas formais explica a maior parcela negativa da variação do produto. A comparação com a trajetória brasileira de PTF e os exercícios de bem-estar sugerem uma diferença quantitativa importante entre reformas moderadas e o teto de 36 horas. As magnitudes dependem da fadiga e das horas iniciais: preservar a cauda habitual acima de 44h reduz o requisito em 40 horas para cerca de $0{,}07\%$. Os resultados motivam a avaliação de trajetórias graduais e de instrumentos de transição, sem constituir uma estimativa causal dos efeitos de uma proposta legislativa específica.

\vspace{0.05cm}
\noindent{\small \textbf{Palavras-chave:} jornada de trabalho; informalidade; produtividade; Brasil.

\noindent \textbf{Classifica\c{c}\~ao JEL:} E24; J22; J46.}

\newpage
\end{abstract}
\renewcommand{\abstractname}{Abstract}
\begin{center}
\textbf{Working hours, productivity, and informality: a quantitative assessment for Brazil}
\end{center}
\begin{abstract}
\noindent This paper quantifies the total factor productivity (TFP) gain required to offset the short-run output loss associated with alternative weekly working-hour ceilings in Brazil. We combine moments from the 2024Q4 PNAD Contínua with a static partial-equilibrium model in which formal and informal labor are imperfect substitutes, hourly efficiency depends on working hours, and formalization wedges govern the composition of employment. To represent reductions from 44 to 40 and 36 hours, we initially cap formal workers' usual hours at 44 as a baseline measurement assumption. The 40-hour alternative requires $1.53$--$1.57\%$ higher TFP. Moving to 36 hours requires $6.52$--$8.38\%$, depending on the efficiency assumption below 40 hours; informality rises by $2.28$--$2.92$ percentage points in the same counterfactual. Decompositions show that fewer physical formal hours account for the largest negative component of the output change. Comparison with Brazil's historical TFP path and welfare exercises suggests an important quantitative difference between moderate reforms and the 36-hour ceiling. Magnitudes depend on fatigue and baseline hours: retaining the observed usual-hours tail above 44 reduces the requirement at 40 hours to approximately $0.07\%$. The results motivate the assessment of gradual paths and transition instruments, without providing a causal estimate of the effects of a specific legislative proposal.

\end{abstract}

\FloatBarrier

\newpage
 \section{Introdu\c{c}\~ao}\label{sec:intro}

Reformas da jornada legal alteram simultaneamente a utiliza\c{c}\~ao do trabalho, a produtividade por hora e os incentivos de formaliza\c{c}\~ao. Essa intera\c{c}\~ao \'e especialmente relevante no Brasil, onde grande parcela do emprego n\~ao est\'a diretamente coberta pelas restri\c{c}\~oes aplic\'aveis ao setor formal. O debate legislativo de 2025 e abril de 2026 tamb\'em deixou de se resumir a uma escolha bin\'aria entre 44 e 36 horas: passaram a coexistir propostas de 40 horas, escala 5x2 e trajet\'orias faseadas \citep{CamaraPEC82025Protocol,CamaraFortyHours2025,CamaraFortyHoursBill2026}. A quest\~ao quantitativa deste artigo \'e: qual ganho de PTF manteria o produto agregado constante em cada ponto dessa trajet\'oria?

Para respond\^e-la, constru\'imos um \'indice de produtividade requerida, $\Areq$: a varia\c{c}\~ao \'unica de PTF que recomp\~oe exatamente o produto de curto prazo ap\'os a redu\c{c}\~ao do teto. O objeto n\~ao \'e uma previs\~ao de produtividade, mas um contrafactual de neutralidade que permite comparar desenhos de reforma e coloc\'a-los na escala da experi\^encia macroecon\^omica brasileira.

Nosso modelo combina quatro ingredientes. Primeiro, um agregador CES incorpora substitui\c{c}\~ao imperfeita entre trabalho efetivo formal e informal. Segundo, a produtividade hor\'aria varia com a jornada, com pico de efici\^encia assumido em 40 horas. Terceiro, cunhas e custos de ajuste governam a composi\c{c}\~ao formal--informal, com uma extens\~ao setorial. Quarto, prefer\^encias Greenwood--Hercowitz--Huffman (GHH) fornecem um diagn\'ostico transparente de consumo e lazer. Com a elasticidade de substitui\c{c}\~ao fixada em $1{,}326$, a raz\~ao de remunera\c{c}\~ao hor\'aria da PNAD Cont\'inua de 2024T4 disciplina o peso tecnol\'ogico formal. Os resultados tamb\'em cobrem sensibilidades de par\^ametros.

Os resultados revelam uma n\~ao linearidade substantiva. A base limita previamente as horas habituais formais da PNAD a 44h, como aproxima\c{c}\~ao para avaliar as redu\c{c}\~oes do teto, sem trat\'a-las como horas contratadas observadas. A alternativa de 40 horas requer $1{,}53$--$1{,}57\%$ de PTF. No teto de 36 horas, $\Areq=6{,}52\%$ quando a fun\c{c}\~ao de efici\^encia penaliza apenas jornadas acima de 40 horas e $8{,}38\%$ quando a penalidade \'e bilateral. A informalidade aumenta entre $2{,}28$ e $2{,}92$ pontos percentuais, mas a decomposi\c{c}\~ao mostra que a redu\c{c}\~ao f\'isica das horas formais \'e a maior parcela negativa da varia\c{c}\~ao do produto. A magnitude depende da efici\^encia e da distribui\c{c}\~ao inicial: preservar a cauda habitual acima de 44h reduz o requisito para 40 horas a cerca de $0{,}07\%$.

O artigo contribui em tr\^es dimens\~oes. Em primeiro lugar, aproxima a
literatura sobre regula\c{c}\~ao do tempo de trabalho da evid\^encia brasileira sobre jornada, emprego e
produtividade. \citet{Fernandes1991} mostrou que os efeitos de uma redu\c{c}\~ao
generalizada da jornada n\~ao podem ser inferidos diretamente da resposta de
uma firma isolada, enquanto \citet{GonzagaMenezesFilhoCamargo2003} estudam a
redu\c{c}\~ao constitucional de 48 para 44 horas e documentam seus efeitos
sobre jornada efetiva, emprego e sal\'arios. Em uma perspectiva agregada,
\citet{BarbosaFilhoPessoa2014} mostram que medir o insumo trabalho pelo total
de horas, em vez de apenas pelo n\'umero de pessoas ocupadas, altera
substantivamente a leitura da produtividade brasileira. Este artigo
complementa esses trabalhos ao construir um contrafactual prospectivo que
relaciona diferentes tetos legais \`a efici\^encia hor\'aria e ao ganho de PTF
necess\'ario para preservar o produto.

Em segundo lugar, o artigo incorpora explicitamente a margem
formal--informal. A literatura brasileira mostra que formalidade e
informalidade n\~ao devem ser interpretadas como estados isolados ou como uma
simples diferen\c{c}a contratual. Os diferenciais observados de remunera\c{c}\~ao
refletem sele\c{c}\~ao dos trabalhadores
\citep{MenezesFilhoMendesAlmeida2004}, e as transi\c{c}\~oes entre emprego
formal, informalidade, trabalho por conta pr\'opria e desemprego s\~ao
quantitativamente relevantes \citep{CuriMenezesFilho2006,HirataMachado2010}.
Ao mesmo tempo, institui\c{c}\~oes trabalhistas e mecanismos de
fiscaliza\c{c}\~ao afetam conjuntamente formaliza\c{c}\~ao, desemprego e
bem-estar \citep{Ulyssea2008,Ulyssea2018}, enquanto a din\^amica do emprego
formal tamb\'em est\'a associada ao tamanho dos estabelecimentos
\citep{CorseuilMouraRamos2011}. O modelo incorpora essa margem por meio da
substitui\c{c}\~ao imperfeita entre trabalho formal e informal e de cunhas de
formaliza\c{c}\~ao. Isso permite avaliar n\~ao apenas o efeito agregado da reforma, mas tamb\'em como as respostas de produto e composi\c{c}\~ao do emprego diferem entre setores.

Em terceiro lugar, o artigo organiza os contrafactuais em uma curva completa
de $\Areq$. Em vez de avaliar somente um ponto final, o exerc\'icio acompanha
toda a trajet\'oria entre 44 e 36 horas e identifica a n\~ao linearidade
associada \`a efici\^encia hor\'aria. Essa representa\c{c}\~ao permite
distinguir reformas que predominantemente eliminam horas associadas \`a fadiga
daquelas que passam a comprimir horas tratadas como efetivas pelo modelo. A
decomposi\c{c}\~ao entre perda mec\^anica de horas, resposta da efici\^encia e
realoca\c{c}\~ao formal--informal fornece ainda uma interpreta\c{c}\~ao
econ\^omica dos resultados: a redu\c{c}\~ao de horas formais domina as parcelas negativas da resposta do produto agregado; a resposta da efici\^encia depende da especifica\c{c}\~ao, e a realoca\c{c}\~ao para a informalidade amplia a perda.

A op\c{c}\~ao por um benchmark est\'atico de curto prazo, com capital
predeterminado e emprego agregado mantido no n\'ivel inicial, torna transparente
a interpreta\c{c}\~ao de $\Areq$: trata-se da produtividade necess\'aria para
compensar a redu\c{c}\~ao de horas antes que outros mecanismos de ajuste tenham
se completado. As reformas do Brasil em 1988 e de Portugal em 1996 fornecem
contexto externo para interpretar as margens de resposta
\citep{GonzagaMenezesFilhoCamargo2003,AsaiLopesTondini2024}. Em conjunto com
as sensibilidades de par\^ametros, essas compara\c{c}\~oes permitem apresentar os
resultados como refer\^encias quantitativas condicionais para o desenho e o
sequenciamento de reformas da jornada no Brasil.

O restante do artigo est\'a organizado da seguinte forma. A Se\c{c}\~ao~\ref{sec:facts} apresenta os fatos que disciplinam o exerc\'icio. A Se\c{c}\~ao~\ref{sec:model} descreve o modelo. A Se\c{c}\~ao~\ref{sec:calibration} discute calibra\c{c}\~ao e identifica\c{c}\~ao; a Se\c{c}\~ao~\ref{sec:validation} apresenta checagens externas. A Se\c{c}\~ao~\ref{sec:results} reporta os contrafactuais e a Se\c{c}\~ao~\ref{sec:policy} examina implica\c{c}\~oes de transi\c{c}\~ao. A Se\c{c}\~ao~\ref{sec:conclusion} conclui.

 \section{Ambiente institucional e fatos brasileiros}\label{sec:facts}

\subsection{A reforma e os contrafactuais relevantes}

A PEC 8/2025 foi apresentada com uma semana de quatro dias e teto de 36 horas \citep{CamaraPEC82025Protocol}. O debate de 2025 e abril de 2026 passou a incluir teto de 40 horas, escala 5x2 e transições faseadas; em 22 de abril de 2026, a admissibilidade da PEC 8/2025 e da PEC 221/2019 foi aprovada na Comissão de Constituição e Justiça \citep{CamaraFortyHours2025,CamaraFortyHoursBill2026,CamaraAdmissibility2026,CamaraSpecialCommission2026}. A análise, portanto, considera toda a sequência de tetos entre 44 e 36 horas, em vez de tratar apenas o ponto final de maior intensidade.

\subsection{Horas formais, informalidade e produtividade}

As horas formais estão concentradas em jornadas de 40 horas ou mais, a margem restringida por um teto de 36h. A PNAD Contínua de 2024T4 informa média habitual de $41{,}42$ horas entre formais e $35{,}48$ entre informais no trabalho principal \citep{PNAD2024T4Microdados}. São horas habituais, distintas das efetivas e das contratadas. Para representar a base de 44h, limitamos previamente a distribuição formal a esse teto, reduzindo sua média para $39{,}95$ horas; a jornada informal permanece na média observada. Preservar a cauda acima de 44h é uma sensibilidade separada.\footnote{Na PNAD, $17{,}97\%$ dos formais reportam mais de 44 horas habituais. A limitação inicial é uma hipótese de mensuração e exposição à reforma, não uma observação de contratos. A formalidade estatística não equivale à cobertura de uma mesma regra de jornada; o exercício não identifica sua aplicação a contratos e horas extras.}

A informalidade é grande e dispersa. A taxa nacional é de $38{,}60\%$, combinando posição na ocupação e registro do negócio no CNPJ \citep{PNADDicionario2022}. A faixa setorial vai de $34{,}21\%$ em serviços a $75{,}75\%$ na agricultura (Tabela~\ref{tab:facts_informality}). A remuneração formal--informal também disciplina o problema: a razão de massas de renda habitual por totais de horas é de $1{,}62244$ entre ocupados remunerados. Esse momento inclui conta própria e empregadores e não é um prêmio salarial condicional. Ao limitar também as horas formais no denominador da ponte a 44h, a razão passa a $1{,}68247$. A calibração usa essa razão compatível com a base para ajustar o peso tecnológico formal, mantendo a elasticidade CES de referência.

A PTF brasileira permaneceu estagnada ou em queda ao longo do período pós-1990. A série \texttt{rtfpna} da Penn World Table 11.0 mostra queda anualizada de $-0{,}83\%$ ao ano de 1990 a 2023 e $-0{,}82\%$ no horizonte pré-Covid de 1990 a 2019 \citep{FeenstraInklaarTimmer2015,PWT2025,FREDPWTBrazil2026}. Entre janelas de dez anos inteiramente contidas em 1990--2019, a melhor década móvel é de 2000 a 2010, com apenas $+0{,}05\%$ ao ano. O benchmark de PTF na Seção~\ref{sec:results} usa essa série atualizada como referência de escala para $A_{\text{req}}$.

Em conjunto, esses fatos disciplinam três margens: a exposição dos trabalhadores formais ao novo teto, a composição inicial formal--informal e a escala do ganho de produtividade requerido. A Tabela~\ref{tab:facts_informality} documenta a heterogeneidade setorial que motiva a extensão do exercício.\footnote{A calibração nacional utiliza um grupo Brasil e a extensão desagrega a economia por setor. As participações por porte da RAIS referem-se a vínculos formais em estabelecimentos, sem uma ligação validada com o universo de pessoas da PNAD \citep{RAIS2022TabelasVerificadas}. As comparações entre fontes e as definições de dados são documentadas no apêndice online.}

\begin{table}[t]
\centering
\caption{Emprego, informalidade e horas habituais, PNAD Contínua 2024T4.}\label{tab:facts_informality}
\small
\begin{tabular}{lrrrr}
\toprule
 & Parcela do & Informalidade & \multicolumn{2}{c}{Horas semanais médias} \\
\cmidrule(lr){4-5}
Setor & emprego (\%) & (\%) & Formais & Informais \\
\midrule
Agricultura & 7{,}574 & 75{,}750 & 45{,}206 & 35{,}680 \\
Indústria e construção & 20{,}474 & 40{,}258 & 42{,}513 & 37{,}081 \\
Serviços & 71{,}934 & 34{,}211 & 40{,}996 & 34{,}904 \\
Não classificado & 0{,}019 & 60{,}027 & 42{,}111 & 36{,}902 \\
\midrule
Brasil & 100{,}000 & 38{,}600 & 41{,}424 & 35{,}485 \\
\bottomrule
\end{tabular}
\begin{minipage}{0.96\textwidth}
\footnotesize\justifying
\textit{Notas:} Ocupados de 14 anos ou mais; trabalho principal; peso V1028; amostra de 209.311 pessoas. Informalidade combina VD4009 com o indicador de CNPJ V4019. Agricultura corresponde às divisões 01--03 da CNAE Domiciliar 2.0; indústria e construção, a 05--43; serviços, a 45--99. Os casos não classificados são preservados na estatística nacional e excluídos explicitamente na extensão de três setores, com renormalização das parcelas. A tabela apresenta as horas observadas, antes da limitação formal a 44h usada no modelo. São estimativas pontuais ponderadas, sem erros-padrão de desenho amostral; diferenças de soma decorrem de arredondamento. \textit{Fonte:} microdados do IBGE \citep{PNAD2024T4Microdados}.
\end{minipage}
\end{table}
 \section{Modelo}\label{sec:model}

O modelo traduz os fatos da Seção~\ref{sec:facts} em uma contabilidade macroeconômica de curto prazo. Há um grupo nacional de firmas e dois tipos de trabalho, formal e informal; a extensão setorial usa grupos indexados por $g$. O emprego total de cada grupo é fixo no horizonte do contrafactual; firmas escolhem sua composição formal--informal, enquanto capital e produtividade agregada são inicialmente predeterminados.

\subsection{Tecnologia e eficiência das horas}

Uma firma representativa do grupo $g$ produz $Y_g$ combinando capital $K_g$ e trabalho efetivo $L_g$ por uma tecnologia Cobb--Douglas:
\begin{equation}\label{eq:production}
Y_g=A_g K_g^\alpha L_g^{1-\alpha},\qquad 0<\alpha<1,
\end{equation}
em que $A_g$ denota a PTF e $\alpha$ é a elasticidade do produto em relação ao capital. O insumo trabalho é um agregado CES de trabalho efetivo formal e informal:
\begin{equation}\label{eq:ces}
L_g=\left[\omega L_{F,g}^{\rho}+(1-\omega)L_{I,g}^{\rho}\right]^{1/\rho},
\qquad \rho=\frac{\sigma_{\mathrm{sub}}-1}{\sigma_{\mathrm{sub}}}.
\end{equation}
em que $\omega$ é o peso tecnológico do trabalho formal e $\sigma_{\mathrm{sub}}$ é a elasticidade de substituição. Os insumos efetivos são
\begin{align}
L_{F,g}&=N_{F,g}q_{F,g}(\bar h), &
q_{F,g}(\bar h)&=\sum_b\theta_{b,g}\,h_{b,g}(\bar h)e\bigl(h_{b,g}(\bar h)\bigr),\label{eq:effective_formal}\\
L_{I,g}&=\eta_I N_{I,g}h_{I,g}e(h_{I,g}), &
h_{b,g}(\bar h)&=\min\{h_{b,g}^{0},\bar h\}.
\end{align}
O emprego total satisfaz $N_{I,g}=N_g-N_{F,g}$. Entre os formais, $b$ indexa as faixas observadas de horas habituais, com participações $\theta_{b,g}$. Para representar a redução a partir do teto de 44h, usamos $h_{b,g}^{0}=\min\{h_{b,g}^{\mathrm{obs}},44\}$ no estado inicial e $\bar h\leq44$ no contrafactual. Essa limitação é uma hipótese de representação da jornada, não uma observação de horas contratuais. No setor informal, mantemos a média observada $h_{I,g}$. O parâmetro $\eta_I$ mede a eficiência relativa do insumo informal. A redistribuição de trabalhadores preserva a distribuição de horas dentro de cada modalidade.

A eficiência horária acima do pico $h^*=40$ segue $e(h)=\exp\{-\kappa(h-h^*)^2\}$. A evidência de \citet{Pencavel2015} motiva a perda de eficiência associada a jornadas longas, mas não identifica a curva brasileira nem o formato abaixo do pico. Por isso consideramos duas especificações:
\begin{equation}\label{eq:efficiency_modes}
e_B(h)=\exp\{-\kappa(h-h^*)^2\},\qquad
e_F(h)=\exp\{-\kappa[\max\{h-h^*,0\}]^2\}.
\end{equation}
Na especificação de \emph{fadiga apenas acima de 40 horas}, $e_F(h)=1$ para $h\leq h^*$. Na alternativa de \emph{penalidade bilateral}, a mesma forma quadrática vale dos dois lados de $h^*$, penalizando também jornadas menores. O pico refere-se à eficiência por hora, não necessariamente ao trabalho efetivo total $he(h)$.

\subsection{Formalização e decisão da firma}

Tomando o emprego total como dado, a firma escolhe continuamente $N_{F,g}\in[0,N_g]$ para maximizar
\begin{equation}\label{eq:firm}
J_g=Y_g-\tau_g N_{F,g}-\frac{\pi_g}{2}N_{I,g}^2
-\frac{\gamma_g}{2}(N_{F,g}-N_{F,g}^0)^2.
\end{equation}
O custo linear $\tau_g$ resume obrigações formais; $\pi_g$ governa o custo convexo da informalidade; e $\gamma_g$ penaliza desvios em relação ao emprego formal pré-reforma $N_{F,g}^0$. As cunhas representam distorções privadas, no espírito de \citet{HsiehKlenow2009} e \citet{RestucciaRogerson2008}. \citet{DixCarneiroGoldbergMeghirUlyssea2026} fornecem uma microfundamentação relacionada para formalidade endógena. A redução do teto altera $L_{F,g}$ e, portanto, a condição marginal que determina a composição do emprego.

Definindo $MP_{F,g}=\partial Y_g/\partial N_{F,g}$ e $MP_{I,g}=\partial Y_g/\partial N_{I,g}$ com a outra modalidade fixa, a condição interior é
\begin{equation}\label{eq:firm_foc}
J_g'=MP_{F,g}-MP_{I,g}-\tau_g+\pi_g N_{I,g}
-\gamma_g(N_{F,g}-N_{F,g}^0)=0.
\end{equation}
Verificam-se também as fronteiras: $J_g'(0+)\leq0$ na inferior e $J_g'(N_g-)\geq0$ na superior.

No estado inicial, um alvo de informalidade identifica somente $\tau_g-\pi_gN_{I,g}^0=D_g$, em que $D_g=MP_{F,g}^0-MP_{I,g}^0$. Para separar as cunhas impomos
\begin{equation}\label{eq:wedge_normalization}
\tau_g\geq0,\qquad\pi_g\geq0,\qquad\tau_g\pi_g=0.
\end{equation}
As cunhas são recalibradas sob essa normalização em cada especificação, sem identificação independente de ambas. O apêndice apresenta a solução completa.

\subsection{Produtividade requerida e bem-estar}

Definimos $\Areq$ como a variação de PTF que restaura o produto agregado. Seja $N_{F,g}^*(a,\bar h)$ a escolha da firma quando a produtividade é multiplicada por $a>0$:
\begin{equation}\label{eq:areq}
\sum_gY_g\!\left(a A_g,K_g,\bar h,N_{F,g}^*(a,\bar h)\right)=Y_0,
\qquad \Areq=a-1.
\end{equation}
Essa definição transforma a perda de produto em uma unidade comparável entre tetos e com a história da PTF. O produto de referência $Y_0$ é o da base de 44h, de modo que $\Areq(44)=0$. Ela não impõe nem prevê que a reforma produza tal ganho. A busca reotimiza a composição em cada avaliação de $a$; $A_{\mathrm{req}}^{\mathrm{fixo}}$ mantém $N_{F,g}=N_{F,g}^0$. Ambas restauram produto bruto, com requisito assinado: se negativo, não há necessidade de ganho adicional.

A decomposição atualiza sucessivamente \emph{horas físicas, eficiência e realocação}, gerando os níveis $Y_H$, $Y_E$ e $Y_1$. Em cada passo preservam-se os componentes ainda não alterados do estado inicial:
\begin{equation}\label{eq:decomposition}
\frac{Y_1-Y_0}{Y_0}
=\frac{Y_H-Y_0}{Y_0}+\frac{Y_E-Y_H}{Y_0}+\frac{Y_1-Y_E}{Y_0}.
\end{equation}
O denominador comum garante a soma das parcelas; a atribuição das interações depende da ordem declarada.

Para o diagnóstico de bem-estar, adotamos preferências GHH \citep{GreenwoodHercowitzHuffman1988}, com elasticidade de Frisch $1/\nu=0{,}5$. A restrição de recursos é $\mathcal C+\mathcal J=Y$, em que $\mathcal C$ é o consumo agregado e $\mathcal J=\sum_g\gamma_g(N_{F,g}-N_{F,g}^0)^2/2$ é o custo real de ajuste. Os pagamentos associados a $\tau_g$ e $\pi_g$ são transferências devolvidas ao domicílio representativo; a alternativa de custos reais consta do apêndice.

O composto GHH mede consumo líquido da desutilidade monetizada do trabalho. Com consumo por trabalhador $C=\mathcal C/\sum_gN_g$, média de horas físicas $h$ e $G=C-v(h)$, em que $v(h)=\psi h^{1+\nu}/(1+\nu)$, reportamos sua variação e o equivalente de consumo:
\begin{equation}\label{eq:cv}
\Delta GHH=\frac{G_1-G_0}{G_0},\qquad
CE=\frac{C_1-C_0-v(h_1)+v(h_0)}{C_0}.
\end{equation}
Os resultados percentuais multiplicam essas expressões por 100. O indicador não incorpora saúde, produção doméstica, barganha, transições de emprego ou pesos distributivos e, portanto, não deve ser interpretado como uma ordenação completa de bem-estar social.

\subsection{Fechamento de curto prazo}

O fechamento é estático e de equilíbrio parcial. Capital e emprego total são predeterminados; retroalimentações salário--preço, monetárias, fiscais e intersetoriais permanecem fora do modelo. O apêndice online organiza os canais incluídos, delimitados e omitidos. O exercício não quantifica ajuste endógeno de capital nem incidência distributiva, pois não inclui equações de investimento ou alocação individual de consumo.
 \section{Calibração e disciplina empírica}\label{sec:calibration}

\subsection{Parâmetros e momentos-alvo}

A Tabela~\ref{tab:params} lista os parâmetros. As entradas observadas vêm da PNAD Contínua \citep{PNAD2024T4Microdados}; parâmetros calibrados por momentos ($\omega$, $\tau_g$, $\pi_g$, $\psi$) são ajustados internamente. Parâmetros escolhidos pelo autor descrevem margens motivadas pela evidência sobre horas de trabalho em \citet{Pencavel2015} e \citet{CollewetSauermann2017}, a literatura de produtividade informal \citep{LaPortaShleifer2014} e a evidência transnacional sobre horas em \citet{BickFuchsSchundelnLagakos2018}, sem serem identificados por esses estudos para o Brasil. O apêndice online apresenta fontes e sensibilidade completas para cada parâmetro.

\begin{table}[t]
\centering
\caption{Valores de parâmetros por base evidencial.}\label{tab:params}
\small
\begin{tabular}{@{}p{2.1cm}p{4.1cm}p{8.1cm}@{}}
\toprule
Parâmetro & Valor & Base da escolha \\
\midrule
\multicolumn{3}{l}{\textit{Entradas PNAD 2024T4 e regra de base de 44h}} \\
$i_0$ & $0{,}3860009$ & Informalidade, VD4009 e V4019. \\
$(h_b^0,\theta_b)$ & Base limitada a 44h & Horas habituais formais com $h_b^0=\min\{h_b^{obs},44\}$. \\
$h_I$ & $35{,}48475$ & Média habitual informal no trabalho principal. \\
$R_h$ & $1{,}682473$ & Renda por horas; denominador formal limitado a 44h. \\
\midrule
\multicolumn{3}{l}{\textit{Calibração condicional: bilateral / fadiga acima do pico}} \\
$\omega$ & $0{,}675651$ / $0{,}676650$ & Ajuste de $R_h$ com $\sigma_{\mathrm{sub}}$ e $\eta_I$ fixos. \\
$\tau$ & $2{,}661015$ / $2{,}714237$ & Condição inicial de composição e normalização~\eqref{eq:wedge_normalization}. \\
$\pi$ & $0$ / $0$ & Mesma normalização; não identificado separadamente. \\
$\kappa$ & $0{,}002109804$ & Derivado das hipóteses $E_Q$, $h^*$ e $h_{\mathrm{ref}}$. \\
$\psi$ & $(8{,}2548;\,8{,}4199)\times10^{-5}$ & Normalização da condição representativa de horas. \\
\midrule
\multicolumn{3}{l}{\textit{Hipóteses e unidades mantidas}} \\
$\sigma_{\mathrm{sub}}$ & $1{,}326$ & Elasticidade CES de referência; sensibilidade no apêndice. \\
$\alpha;\,\eta_I$ & $0{,}35;\,0{,}40$ & Elasticidade do capital e eficiência relativa informal. \\
$E_Q;\,h^*$ & $0{,}60;\,40$h & Elasticidade local de $he(h)$ e pico de eficiência horária. \\
$h_{\mathrm{ref}}$ & $42{,}244$h & Âncora externa da eficiência, separada da nova média. \\
$\gamma;\,\nu$ & $0{,}06;\,2$ & Custo de ajuste; inverso da elasticidade de Frisch. \\
$N;\,K;\,A$ & $0{,}59;\,1;\,1$ & Unidades do modelo; $N$ e $K$ fixos. \\
\bottomrule
\end{tabular}
\begin{minipage}{0.97\textwidth}
\footnotesize\justifying
\textit{Notas:} Cunhas em unidades do modelo, não alíquotas observadas. As duas colunas de valores calibrados referem-se às funções da equação~\eqref{eq:efficiency_modes}. $N=0{,}59$ normaliza as unidades dos custos, sem estimar ocupados/população. A distribuição inicial aplica o teto de 44h às horas formais, preservando os pesos PNAD; a média informal permanece observada. \textit{Fonte:} microdados do IBGE e calibração condicional.
\end{minipage}
\end{table}

O parâmetro $\kappa$ deriva da elasticidade local $E_Q=d\log[he(h)]/d\log h=0{,}60$ na referência de $42{,}244$h, uma hipótese tecnológica distinta da média observada. Trata-se da elasticidade do trabalho efetivo por pessoa. O parâmetro $\psi$ é normalizado pela condição inicial de horas do agente representativo. As derivações constam do apêndice.

\subsection{Elasticidade formal--informal e disciplina da calibração}

A calibração não trata $\sigma_{\mathrm{sub}}$ como uma estimativa estrutural direta. O valor central $\sigma_{\mathrm{sub}}=1{,}326$ é um ponto de referência expositivo, não uma elasticidade identificada. Mantemos esse valor e ajustamos $\omega$ à razão formal--informal de remuneração horária da PNAD. A ponte supõe remuneração igual ao produto marginal bruto; não transporta diretamente o prêmio estimado no modelo de busca de \citet{MeghirNaritaRobin2015}.

A razão divide massas de renda habitual positiva por totais de horas válidas do trabalho principal. Para compatibilizá-la com a base, substituímos as horas formais no denominador por $\min\{h^{obs},44\}$, mantendo renda e horas informais. A razão passa de $1{,}622443$ nos dados integrais a $1{,}682473$ sob essa regra. No modelo, $W_{F,g}=MP_{F,g}$ e $W_{I,g}=MP_{I,g}$ são remunerações semanais por trabalhador. Calculamos esses marginais antes de agregar grupos:
\begin{equation}\label{eq:hourly_bridge}
R_h^{\mathrm{mod}}=
\frac{\displaystyle\sum_gN_{F,g}W_{F,g}/\sum_gN_{F,g}\bar h_{F,g}}
{\displaystyle\sum_gN_{I,g}W_{I,g}/\sum_gN_{I,g}h_{I,g}}.
\end{equation}
A média $\bar h_{F,g}$ é de horas físicas. Essa agregação distingue remuneração por hora de remuneração semanal e de remuneração por unidade de trabalho efetivo.

Fixados $\sigma_{\mathrm{sub}}$, $\eta_I$ e os demais parâmetros, igualamos~\eqref{eq:hourly_bridge} a $1{,}682473$, recalibrando as cunhas a cada tentativa de $\omega$. Obtemos $\omega=0{,}675651$ no bilateral e $0{,}676650$ na fadiga acima do pico, distintos da participação formal de $0{,}613999$. A extensão da ponte aos ocupados sem remuneração requer uma hipótese de seleção; o apêndice documenta os universos e as alternativas. Um único momento não identifica conjuntamente tecnologia e cunhas.

Essa disciplina é importante para o sinal da resposta da informalidade. Na referência, $\sigma_{\mathrm{sub}}>1$ e os insumos são substitutos imperfeitos: no teto de 36 horas, a redução do trabalho efetivo formal desloca empregos marginais para a informalidade. \citet{DerenoncourtGerardLagosMontialoux2025} documentam efeitos do salário mínimo sobre salários informais e realocação limitada para fora do emprego formal no Brasil. Essa evidência situa a interação entre regulação e informalidade; sua elasticidade de realocação não identifica diretamente $\sigma_{\mathrm{sub}}$. O apêndice online reporta as grades completas dos parâmetros CES, com pontos interiores e ponte recalibrada em cada ponto, e discute a distinção entre elasticidades na literatura.

Como robustez, reconstruímos a calibração preservando a distribuição habitual integral, inclusive acima de 44h, e variamos eficiência, pico, distribuição de horas e tratamento dos custos. As sensibilidades incluem o cenário sem fadiga e não constituem um conjunto identificado ou intervalo de confiança.
 \section{Checagens da calibração e benchmarks externos}\label{sec:validation}

O modelo reproduz os momentos-alvo por construção. A validação relevante, portanto, exige momentos não calibrados e comparações externas. Esses exercícios não identificam causalmente o contrafactual brasileiro contemporâneo.

A Tabela~\ref{tab:calibration_fit} reúne momentos-alvo e checagens domésticas. A informalidade inicial e a razão horária são reproduzidas por construção. As horas médias do modelo e a parcela acima de 36h decorrem da distribuição e da regra de limitação a 44h, não constituindo validação comportamental. A razão semanal apresenta diferença de $0{,}01393$ em relação à PNAD. Embora não seja alvo direto, compartilha a renda da ponte horária e depende das médias de horas; sua proximidade é uma checagem descritiva, sem validar as elasticidades. A ponte empírica usa remunerados, enquanto o modelo inclui todos os ocupados.

\begin{table}[t]
\centering
\caption{Alvos, entradas e checagem descritiva no cenário nacional principal.}\label{tab:calibration_fit}
\small
\begin{tabular}{@{}p{6.1cm}rrp{4.4cm}@{}}
\toprule
Momento & Referência & Modelo & Interpretação \\
\midrule
Informalidade inicial (\%) & 38{,}60009 & 38{,}60009 & Alvo das cunhas \\
Razão por hora, base de 44h & 1{,}68247 & 1{,}68247 & Alvo de $\omega$, condicional \\
Horas habituais formais médias & 41{,}42440 & 39{,}94664 & Base limitada a 44h \\
Horas habituais informais médias & 35{,}48475 & 35{,}48475 & Entrada fixa \\
Formais com mais de 36h (\%) & 85{,}58537 & 85{,}58537 & Entrada, identidade de pesos \\
Razão de remuneração semanal & 1{,}88010 & 1{,}89403 & Não alvo; consistência descritiva \\
\bottomrule
\end{tabular}
\begin{minipage}{0.97\textwidth}
\footnotesize\justifying
\textit{Notas:} Os dois modos de eficiência reproduzem os valores do modelo apresentados, até o arredondamento. A referência da razão horária usa as mesmas rendas observadas com as horas formais limitadas a 44h; sem essa transformação, a razão é $1{,}62244$. A referência das horas formais médias é a observada; o modelo aplica a regra de base de 44h. As razões empíricas de remuneração usam renda positiva e horas válidas; o modelo estende a ponte aos ocupados de todas as categorias. Não há regressão com controles individuais. A razão semanal é relacionada algebricamente à razão horária e às horas médias de cada modalidade. Fonte: PNAD Contínua 2024T4 e calibração.
\end{minipage}
\end{table}

As comparações externas consideram a reforma portuguesa de 1996 no teto $44\to40$h e a reforma brasileira de 1988 no teto $48\to44$h. Em Portugal, \citet{AsaiLopesTondini2024} estimam aumento de $4{,}4$ pontos logarítmicos nas vendas por hora nas firmas afetadas. Esse resultado nominal difere da PTF e do produto físico por hora do modelo. A evidência brasileira de \citet{GonzagaMenezesFilhoCamargo2003} trata de horas, remuneração e transições de emprego; a resposta de salário-hora também não equivale à produtividade horária.

Portugal e o Brasil de 1988 diferem do ambiente da PEC 8/2025 em informalidade, arranjo institucional e magnitude da reforma. São referências para margens que o modelo não codifica, não limites empíricos para $\Areq$. Sem harmonizar população e variável de resultado, não permitem classificar o modelo como conservador; o emprego total fixo também impede reproduzir seus testes sobre desemprego. O apêndice apresenta as sensibilidades e as verificações de consistência dos cálculos.
 \section{Resultados quantitativos}\label{sec:results}

\subsection{A curva de produtividade requerida}

\begin{figure}[t]
\centering
\includegraphics[width=0.92\textwidth]{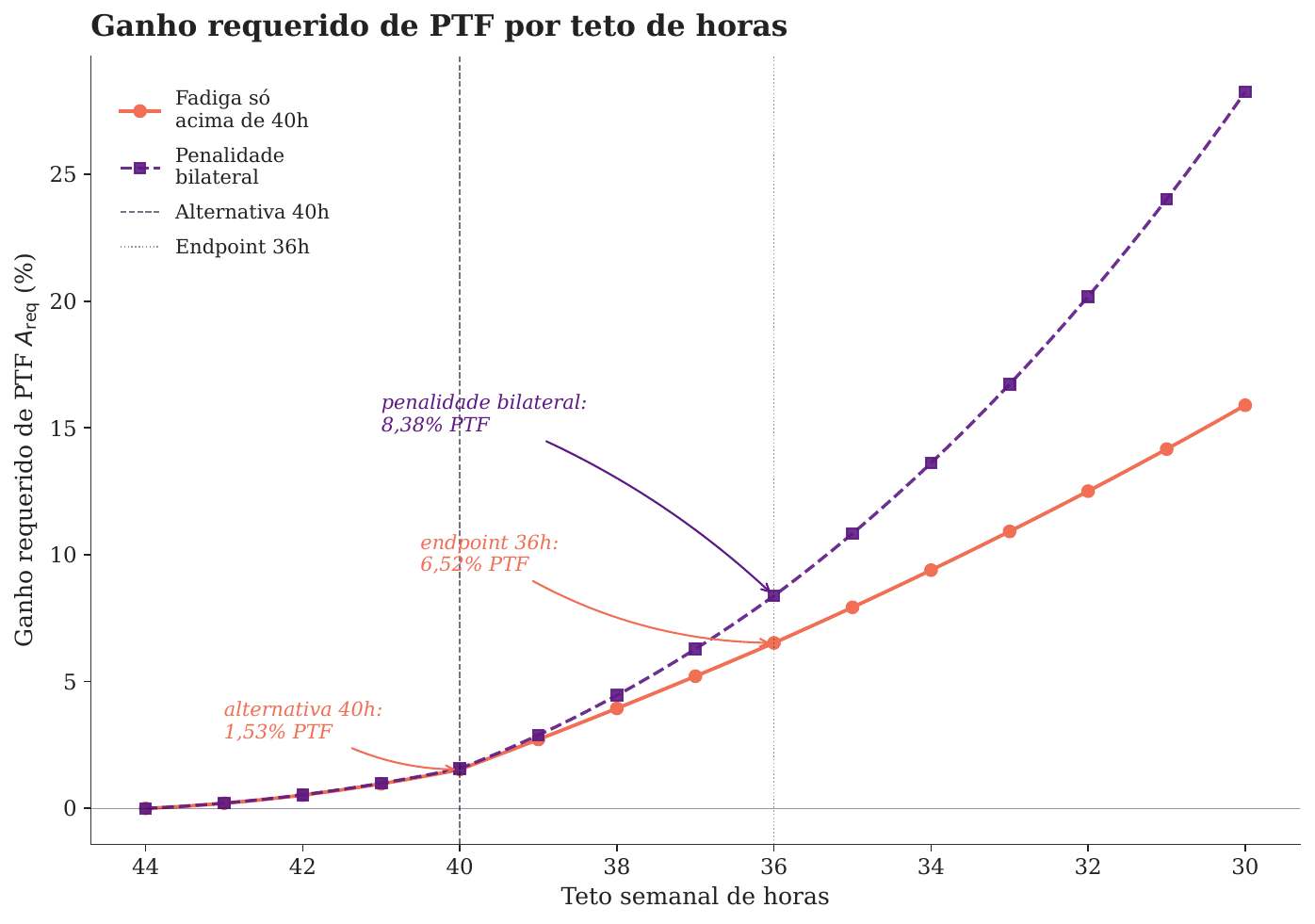}
\caption{Ganho requerido de PTF $\Areq$ como função do teto semanal de horas.}\label{fig:areq_curve}
\tabnotes{A base limita previamente as horas habituais formais a 44h; esse teto é a referência sem reforma. A transformação aproxima a exposição ao limite legal, sem observar horas contratadas. $\Areq$ é assinado e restaura o produto com composição reotimizada. Os marcadores verticais distinguem 40h e 36h. Cada curva tem sua ponte horária e suas cunhas recalibradas. As funções $e(h)$ coincidem acima do pico, mas os agregados podem diferir pela eficiência inicial abaixo dele e pela calibração.}
\end{figure}

O teto de 40 horas e o endpoint de 36 horas são objetos econômicos diferentes. Mover de 44 para 40 horas corta a cauda longa da distribuição formal usada como base e aproxima muitos trabalhadores do pico de eficiência assumido. Mover abaixo de 40 horas já não é sobretudo aparar fadiga; é retirar horas que o modelo trata como efetivas. Por isso a curva da Figura~\ref{fig:areq_curve} dobra para cima.

A Tabela~\ref{tab:mainresults} mostra os tetos de 40 e 36 horas, mas a comparação importante é a inclinação do caminho. A alternativa de 40 horas exige $1{,}53$--$1{,}57\%$ de PTF para restaurar o produto. O endpoint de 36 horas exige uma resposta maior: $6{,}52\%$ sob a hipótese de fadiga só acima de 40h e $8{,}38\%$ sob a penalidade bilateral. Essa diferença não é mero detalhe técnico. No primeiro caso, jornadas mais curtas não recebem penalidade direta de eficiência depois que as horas caem abaixo de 40. No segundo, afastar-se de 40 horas em qualquer direção tem custo de eficiência.

A margem formal--informal se move na direção esperada, mas não é a vilã do resultado agregado. A economia calibrada desloca alguns empregos marginais para a informalidade: em 36 horas, sua participação aumenta $2{,}28$ pontos percentuais sob fadiga só acima do pico e $2{,}92$ pontos sob penalidade bilateral. Ainda assim, a maior parcela negativa da variação do produto vem de menos horas físicas. Sem compensação de PTF, a perda líquida de produto é de $6{,}65\%$ e $8{,}40\%$, respectivamente. Em 40 horas, o produto cai $1{,}64\%$ e $1{,}68\%$.

\begin{table}[!htbp]
\centering
\small
\caption{Resultados nacionais da redução a partir de 44h, com ponte horária recalibrada}\label{tab:mainresults}
\setlength{\tabcolsep}{4pt}
\begin{tabular}{lrrrr}
\toprule
Indicador & Bilateral 40h & Bilateral 36h & Acima 40h & Acima 36h \\
\midrule
$\Delta Y$ (\%) & -1,677 & -8,398 & -1,636 & -6,646 \\
$A_{\mathrm{req}}$ (\%) & 1,573 & 8,381 & 1,534 & 6,524 \\
$A_{\mathrm{req}}$, composição fixa (\%) & 1,494 & 7,956 & 1,457 & 6,194 \\
Informalidade após reforma (\%) & 39,152 & 41,521 & 39,138 & 40,879 \\
$\Delta$ informalidade (p.p.) & 0,552 & 2,921 & 0,538 & 2,279 \\
$\Delta GHH$ (\%) & 0,225 & -4,366 & 0,276 & -2,121 \\
$CE$ (\% de $C_0$) & 0,176 & -3,420 & 0,216 & -1,661 \\
\bottomrule
\end{tabular}
\par\smallskip
\begin{minipage}{0.98\linewidth}\footnotesize Fonte: PNAD Contínua 2024T4 reprocessada e simulações. Acima: fadiga somente acima do pico. Base: horas formais $\min\{h^{obs},44\}$, pesos PNAD preservados, horas informais inalteradas, informalidade de 38,600\%, $\sigma=1{,}326$; $\omega=0{,}675651$ (bilateral) e $0{,}676650$ (acima). A limitação inicial a 44h é uma hipótese de representação, não uma observação de horas contratuais. $A_{\mathrm{req}}$ recompõe o produto dessa base com composição reotimizada em cada avaliação; a variante fixa mantém a composição inicial. $\Delta Y$ usa $Y_0$, $\Delta GHH$ usa $GHH_0$ e $CE$ usa $C_0$ como denominador. Capital e ocupação totais fixos; cunhas são transferências e ajuste consome recursos.\end{minipage}
\end{table}

A figura também ajuda a ler o que o modelo está dizendo. As primeiras quatro horas de redução são baratas em unidades de PTF porque o canal de fadiga trabalha a favor da reforma. As quatro horas seguintes são caras porque a reforma passa a retirar trabalho efetivo. O apêndice mostra como esse diagnóstico depende das horas iniciais: preservar a cauda habitual acima de 44h reduz o requisito em 40 horas para cerca de $0{,}07\%$; em 36 horas, os valores passam a $5{,}00\%$ sob fadiga só acima do pico e $6{,}80\%$ sob penalidade bilateral. Por isso as magnitudes devem ser lidas como resultados da função calibrada de eficiência e da regra de base.
Sem nenhum canal de eficiência das horas, a base de 44h exige $2{,}40\%$ de PTF em 40 horas e $7{,}42\%$ em 36 horas, com ponte e cunhas recalibradas. A perda adicional de eficiência abaixo do pico explica por que, em 36h, a penalidade bilateral exige mais PTF que o cenário sem fadiga.

Congelar a composição formal--informal inicial reduz o requisito em 36 horas para $6{,}19\%$ sob fadiga só acima do pico e $7{,}96\%$ sob penalidade bilateral. A diferença decorre de a composição reotimizada responder ao objetivo privado com cunhas, e não somente ao produto. O apêndice apresenta essa comparação e as demais sensibilidades.

\subsection{Escala macroeconômica e mecanismos}

Ancoramos $\Areq$ no histórico brasileiro de PTF usando a série \texttt{rtfpna} da PWT 11.0 \citep{FeenstraInklaarTimmer2015,PWT2025,FREDPWTBrazil2026}. Essa comparação é uma régua de escala, não uma validação causal do contrafactual, pois o resíduo macro anual da PWT não é o mesmo objeto que a resposta de produtividade de curto prazo do modelo. A PTF brasileira caiu no período pós-1990, então um ganho único de seis e meio a oito e meio por cento é grande nessa régua. A alternativa de 40 horas está em outra região da escala. O apêndice traduz os ganhos únicos em taxas anualizadas para horizontes de três, cinco e dez anos; essa comparação de escala é mais informativa do que perguntar se uma década histórica específica ``repete'' o contrafactual.

\begin{figure}[t]
\centering
\includegraphics[width=0.95\textwidth]{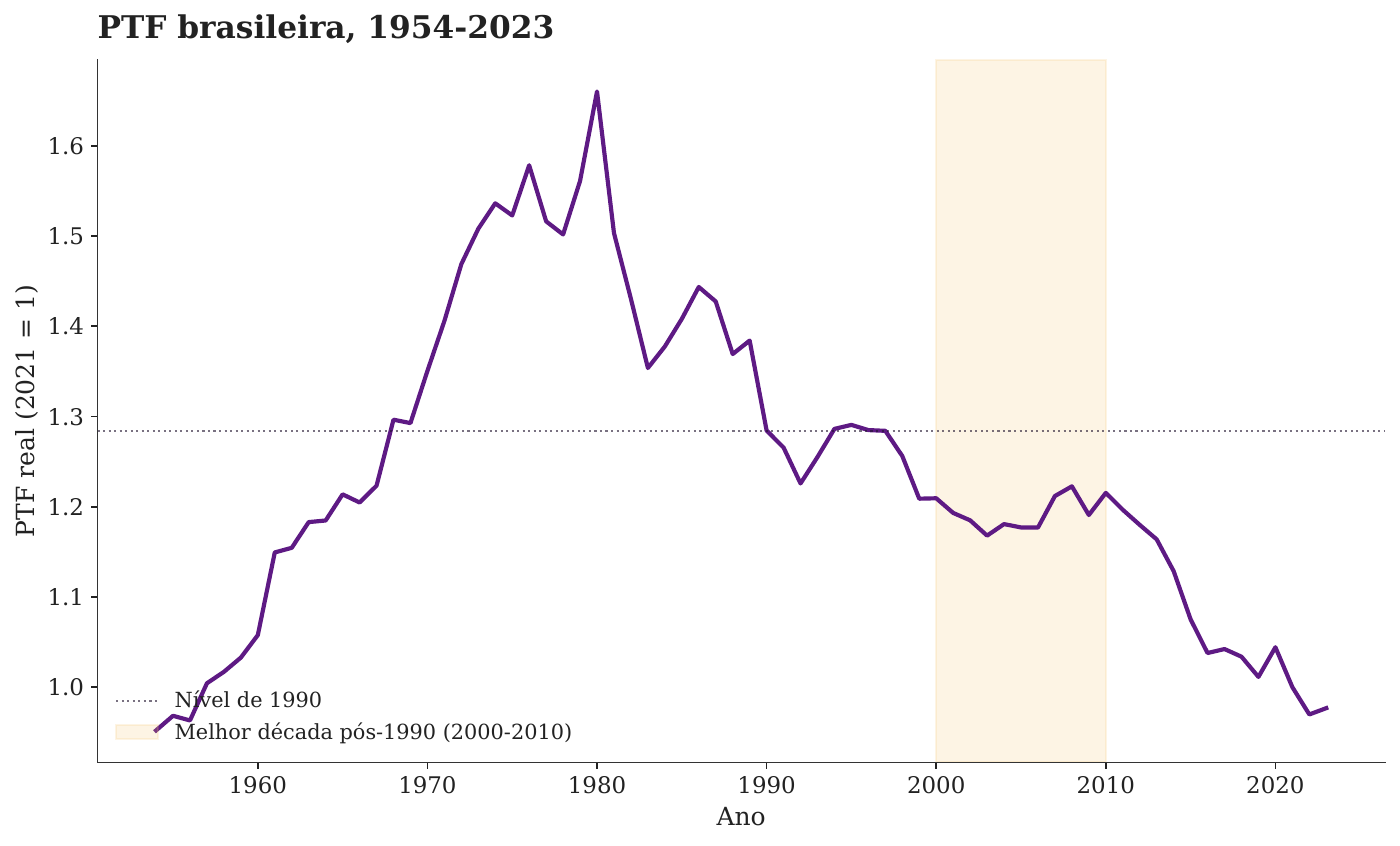}
\caption{PTF brasileira no longo prazo.}\label{fig:tfp_history}
\tabnotes{PTF real a preços nacionais constantes, série \texttt{rtfpna} da PWT 11.0, índice 2021 = 1. A linha horizontal marca o nível de 1990. A faixa destaca 2000--2010, a melhor janela de dez anos inteiramente contida em 1990--2019. Fonte: Penn World Table 11.0 via FRED, série \texttt{RTFPNABRA632NRUG}.}
\end{figure}

Para ilustrar a escala, repartir de forma composta o requisito de 36 horas por cinco anos corresponde a $1{,}27\%$ ao ano sob fadiga só acima do pico e $1{,}62\%$ sob penalidade bilateral. Na série brasileira, a queda anualizada foi de $0{,}83\%$ entre 1990 e 2023.

A decomposição esclarece o mecanismo. Na ordem horas físicas, eficiência e realocação, as parcelas em 36 horas sob penalidade bilateral são $-6{,}7070$, $-0{,}6622$ e $-1{,}0284$ pontos percentuais, totalizando $-8{,}3976\%$. Sob fadiga só acima do pico, a eficiência contribui positivamente, em $0{,}7066$ ponto, enquanto as horas físicas e a realocação contribuem com $-6{,}5397$ e $-0{,}8131$ pontos, totalizando $-6{,}6462\%$. Em 40 horas, sob penalidade bilateral, a perda física de $2{,}244$ pontos é parcialmente compensada pelo ganho de eficiência de $0{,}771$ ponto; a realocação adiciona perda de $0{,}204$ ponto. O apêndice reporta os níveis intermediários, sempre com o mesmo produto inicial como denominador.

\subsection{Bem-estar e heterogeneidade}\label{sec:welfare}

O cálculo de bem-estar deve ser lido como diagnóstico de consumo e lazer. Ele pergunta algo estreito: se emprego é fixo, salários não barganham e a reforma entrega mecanicamente mais tempo fora do trabalho aos incumbentes, o ganho de lazer compensa a perda de consumo? Isso não é um teorema de bem-estar social. Saúde, produção doméstica, barganha, preços, transições de emprego e pesos distributivos ficam fora do cálculo.

Dentro desse objeto estreito, a redução até 40 horas compra lazer a baixo custo de produto. O equivalente de consumo é de $+0{,}18\%$ sob penalidade bilateral e $+0{,}22\%$ sob fadiga só acima do pico; as variações do composto GHH são $+0{,}22\%$ e $+0{,}28\%$. Abaixo de 40 horas, o preço do lazer adicional sobe porque a reforma já não está apenas cortando fadiga de jornadas longas. Em 36 horas, o equivalente de consumo é de $-3{,}42\%$ sob penalidade bilateral e $-1{,}66\%$ sob fadiga só acima do pico; as variações GHH correspondentes são $-4{,}37\%$ e $-2{,}12\%$. As especificações concordam no sinal positivo da redução moderada e no sinal negativo do diagnóstico no endpoint.

\begin{figure}[t]
\centering
\includegraphics[width=0.92\textwidth]{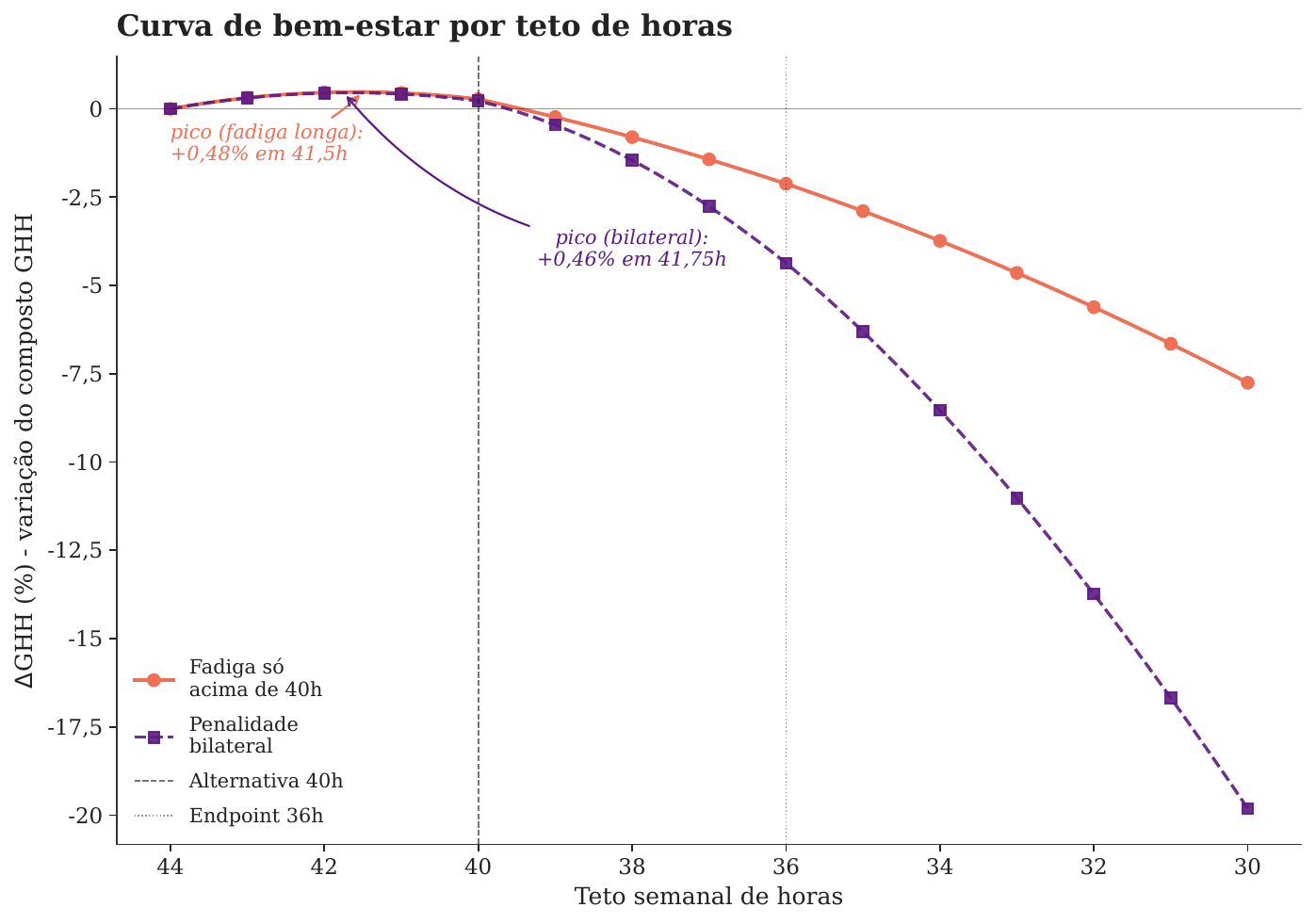}
\caption{Curva de bem-estar por teto de horas sob ambas as calibrações.}\label{fig:welfare_schedule}
\tabnotes{Variação percentual do composto GHH, com denominador $G_0$, conforme a Equação~\eqref{eq:cv}. Distribuição habitual formal limitada inicialmente a 44h, produtividade fixa e composição reotimizada. Os máximos anotados referem-se à grade de 0,25h sob preferências representativas; não identificam jornada socialmente ótima. O equivalente de consumo, com denominador $C_0$, é reportado separadamente na Tabela~\ref{tab:mainresults}.}
\end{figure}

Um ganho de PTF de aproximadamente $1{,}62\%$ no teto de 36h é suficiente para tornar o indicador GHH neutro sob fadiga só acima do pico. O limiar sob penalidade bilateral é de $3{,}40\%$. Ambos ficam abaixo do $\Areq$ neutro em produto porque o lazer compensa parcialmente a perda de consumo dentro desse objeto. O apêndice mostra esses limiares.

A curva do indicador GHH não implica oposição a reduções da jornada. Ela separa a perda de produto sob produtividade fixa, o ganho de produtividade que compensaria essa perda e o diagnóstico de consumo--lazer. Esses objetos não se movem juntos. O sinal do indicador depende da função de eficiência e da distribuição inicial: sem fadiga, o equivalente de consumo é negativo em 40 horas ($-0{,}67\%$) e em 36 horas ($-2{,}52\%$); na sensibilidade com horas habituais integrais, o equivalente de consumo em 36h passa a $-0{,}79\%$ no bilateral e $+0{,}96\%$ na fadiga só acima do pico. O agente representativo não permite decompor sua incidência entre trabalhadores.

Aplicar o arcabouço nacional separadamente a três setores amplos com momentos setoriais da PNAD entrega, no endpoint de 36 horas, um $\Areq$ de $8{,}85\%$ na agricultura, $8{,}97\%$ na indústria e $8{,}16\%$ nos serviços sob penalidade bilateral. Sob fadiga só acima do pico, os valores são $7{,}10\%$, $7{,}12\%$ e $6{,}31\%$. A informalidade sobe em todos os setores nesse teto. Para a alternativa de 40 horas, sob penalidade bilateral, os requisitos são $2{,}09\%$ na agricultura, $1{,}75\%$ na indústria e $1{,}50\%$ nos serviços. Esses são contrafactuais de equilíbrio parcial setor a setor, não uma análise de incidência insumo--produto. O apêndice online reporta os parâmetros, a decomposição completa e as sensibilidades.
 \section{Implicações para o desenho da reforma}\label{sec:policy}

O objeto relevante de política não é apenas a versão original de 36 horas da PEC 8/2025. O debate legislativo de 2025 e abril de 2026 também inclui alternativas de 40 horas, escala 5x2 e transições faseadas. Por isso, o modelo calcula a sequência completa de tetos de horas, e não apenas o endpoint mais intenso.

Um teto de 40 horas é uma reforma moderada nessa aritmética. Na base formal limitada inicialmente a 44h e sob a calibração central de eficiência, ele exige $1{,}53$--$1{,}57\%$ de PTF para restaurar o produto e deixa positivo o indicador GHH representativo. O movimento para 36 horas é maior: afeta $85{,}6\%$ dos trabalhadores formais e exige um salto único de produtividade entre $6{,}52\%$ sob fadiga só acima de 40h e $8{,}38\%$ sob penalidade bilateral. O apêndice mostra como essas magnitudes mudam com a eficiência e a distribuição inicial de horas.

Os requisitos de 36 horas são grandes quando lidos como benchmark de escala contra o histórico brasileiro pós-1990. Na PWT 11.0, a PTF brasileira cai de 1990 a 2023 e também no horizonte pré-Covid de 1990 a 2019. Mesmo a melhor década móvel inteiramente contida em 1990--2019, de 2000 a 2010, cresce apenas $+0{,}05\%$ ao ano. Em um horizonte de cinco anos, o requisito de 36h sob fadiga só acima de 40h equivale a $1{,}27\%$ ao ano de PTF, enquanto o de 40h equivale a $0{,}30\%$ ao ano. Essa comparação não prova inviabilidade; ela apenas coloca o tamanho do ganho requerido na escala da experiência recente brasileira.

A curva é inclinada. O teto de 40 horas mantém o requisito abaixo daquele associado a 36 horas nas especificações examinadas. Mover diretamente para 36 horas impõe uma exigência elevada de produtividade pelo padrão brasileiro pós-1990. Sem esse salto, o modelo produz perda de produto de curto prazo de $6{,}65$ a $8{,}40$ por cento na base principal. O requisito de produtividade é uma referência contábil, não uma previsão.

Uma redução moderada do teto eleva o indicador GHH dentro da calibração central do modelo porque o valor atribuído ao lazer adicional supera o consumo perdido. Em 40 horas, o equivalente de consumo é de $+0{,}18\%$ sob penalidade bilateral e $+0{,}22\%$ sob fadiga só acima do pico. Em 36 horas, ambas as funções produzem perda de bem-estar representativo: $-3{,}42\%$ e $-1{,}66\%$, respectivamente, em equivalente de consumo. A interpretação exige separar produto, PTF requerida e bem-estar: uma redução moderada pode reduzir o produto, elevar o bem-estar representativo e ainda exigir um ganho de PTF positivo. O diagnóstico permanece dependente das hipóteses de fadiga e de mensuração das horas.

O requisito no endpoint varia entre setores. A indústria carrega o maior requisito em 36 horas: $8{,}97\%$ sob eficiência bilateral, próxima da agricultura ($8{,}85\%$) e acima dos serviços ($8{,}16\%$). Todos os três setores veem a informalidade subir. Os serviços concentram a maior parte da perda agregada de produto via sua escala de produto: cerca de $66\%$ a $67\%$ em 36 horas na extensão setorial, conforme a função de eficiência. O exercício setorial é de equilíbrio parcial, setor por setor, e não uma análise insumo-produto de incidência.

A informalidade sobe, mas não explica a maior parte da perda de produto. O modelo calibrado pela razão de remuneração horária formal--informal encontra um aumento agregado de informalidade de $2{,}28$ a $2{,}92$ p.p. em 36 horas. A decomposição mostra que o custo de produto vem sobretudo de menos horas trabalhadas. Ainda assim, a reforma desloca empregos adicionais para a informalidade, enfraquecendo os objetivos formalizadores da política.

\begin{figure}[p]
\centering
\includegraphics[width=0.94\textwidth]{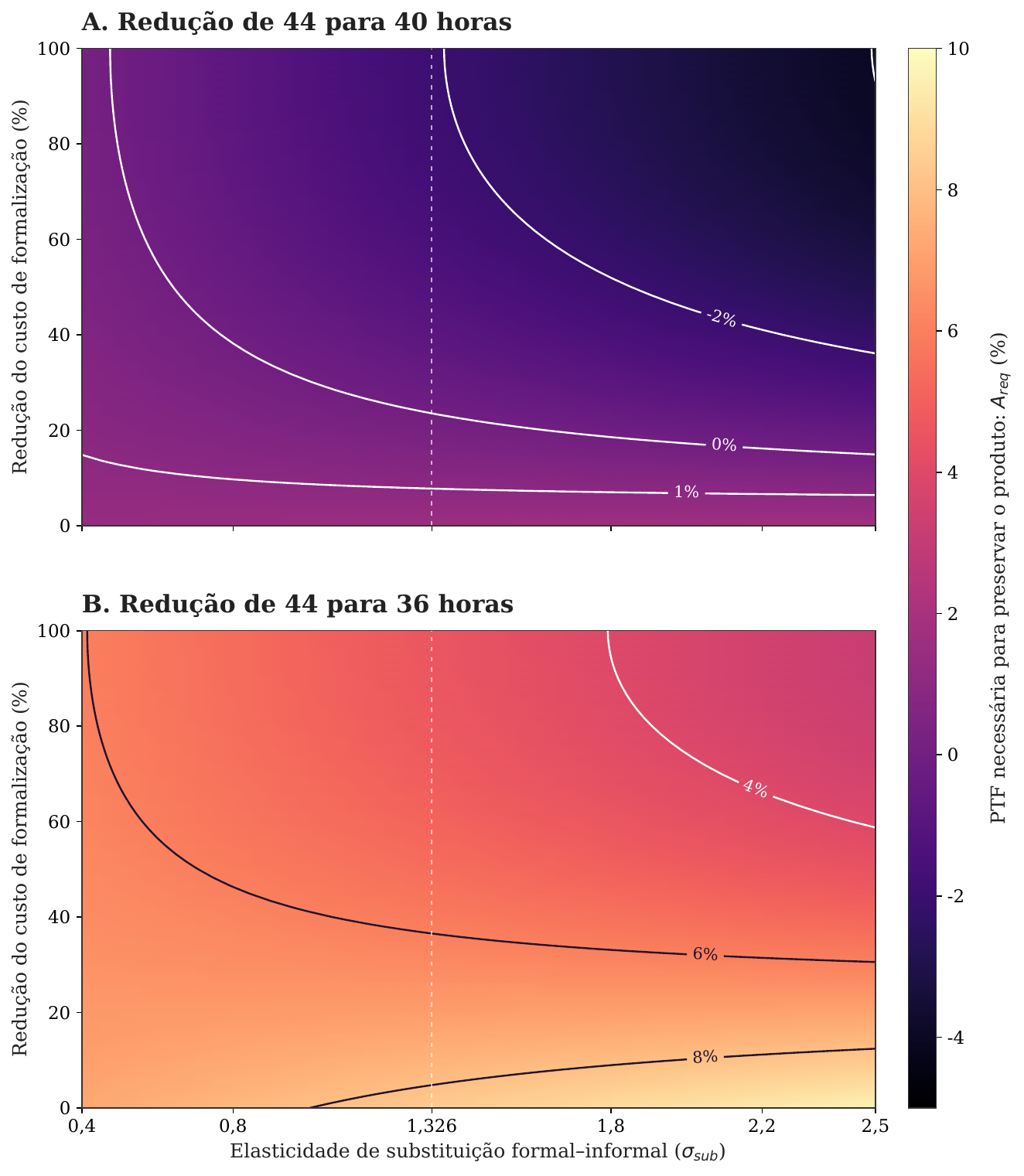}
\caption{Produtividade requerida e redução do custo de formalização: de 44 para 40h (A) e de 44 para 36h (B).}\label{fig:transition_map}
\tabnotes{Ambos os painéis partem da mesma base formal limitada a 44h, com eficiência bilateral e um grupo nacional. O eixo vertical mostra $100r$, em $\tau^{cap}=(1-r)\tau$: $100\%$ elimina o custo adicional $\tau$ do modelo, sem representar uma alíquota fiscal. Para cada $\sigma$, recalibram-se a ponte horária e os custos no estado inicial; depois se reduz somente $\tau$, mantendo $\pi$, capital e âncora de ajuste. Cores e isocurvas usam a mesma escala de $\Areq$, com composição reotimizada em cada avaliação de PTF. Valores negativos correspondem a ganho de produto sob a PTF inicial. A linha vertical tracejada marca $\sigma=1{,}326$. Os painéis são alternativas à base de 44h, não etapas de uma trajetória dinâmica.}
\end{figure}

A Figura~\ref{fig:transition_map} compara os pacotes de transição nas alternativas de 44 para 40 horas (painel A) e de 44 para 36 horas (painel B). Ela cruza a substituição formal--informal, medida por $\sigma$, com a redução do custo de formalização, usando os mesmos eixos e a mesma escala de cores. Esse custo é o componente privado adicional por emprego formal representado por $\tau$ no modelo. No ponto $\sigma=1{,}326$, reduzi-lo em $10\%$ leva o requisito de PTF de $1{,}573\%$ para $0{,}845\%$ em 40h, mas de $8{,}381\%$ para $7{,}606\%$ em 36h.

A diferença de intensidade aparece ao fixar o objetivo de preservar o produto. Em 40h, uma redução de $23{,}53\%$ nesse custo é suficiente para zerar $\Areq$ na calibração central. Em 36h, mesmo eliminar inteiramente esse componente deixa um requisito de PTF de $4{,}62\%$. Portanto, a transição para 36h exige, nesse exercício, ganhos de produtividade além da redução do custo de formalização. O mapa compara custos privados; não identifica o orçamento de um pacote fiscal, sua focalização por porte ou os efeitos de um subsídio temporário.

\paragraph{Por que um pacote de transição importa?} A diferença entre 40 e 36 horas motiva a discussão de instrumentos que facilitem a reorganização do trabalho durante a implementação. Evidência recente de pilotos de semana de quatro dias ajuda a situar essa margem: \citet{FanSchorKellyGu2025} analisam uma intervenção de seis meses em 141 organizações e 2.896 trabalhadores em seis países, com redução de jornada sem corte salarial, e encontram melhora em burnout, satisfação, saúde mental e saúde física. A capacidade de trabalho é autorreportada; o estudo não estima PTF agregada. Esse resultado deve ser lido como evidência de viabilidade organizacional e bem-estar, de modo que complementa, mas não substitui, a aritmética de $\Areq$. A calibração não quantifica o efeito de um pacote fiscal: isso exigiria identificar as cunhas e seu financiamento.

O modelo é de curto prazo e equilíbrio parcial. Capital e emprego total são mantidos fixos, de modo que o modelo desliga investimento e partilha de trabalho. Decisões de investimento e custos de ajuste do capital, discutidos por \citet{Caballero1999}, constituem uma margem adicional para extensões dinâmicas. A evidência sobre partilha de trabalho na Alemanha \citep{Hunt1996} e na reforma constitucional brasileira de 1988 \citep{GonzagaMenezesFilhoCamargo2003}, assim como a experiência portuguesa de 1996 \citep{AsaiLopesTondini2024}, fornece contexto para essas margens, com os limites de comparação discutidos na Seção~\ref{sec:validation}.

Sob esses limites, a aritmética da produtividade motiva a avaliação de caminhos faseados. Uma parada intermediária em 40 horas mantém o requisito de produtividade abaixo do endpoint de 36 horas; o ganho do calendário de transição em si, porém, exige uma análise dinâmica. A implicação é condicional ao modelo: a intensidade da redução e a resposta da produtividade importam para a aritmética de produto e para o indicador de consumo--lazer representativo.

\FloatBarrier
 \section{Conclusão}\label{sec:conclusion}

Este artigo quantificou o ganho de PTF que neutralizaria a perda de produto de curto prazo associada a tetos alternativos de jornada. A curva simulada é acentuadamente não linear: partindo da distribuição habitual formal limitada a 44h e sob a calibração central de eficiência, o teto de 40 horas requer $1{,}53$--$1{,}57\%$ de PTF, enquanto a passagem para 36 horas requer entre $6{,}52\%$ e $8{,}38\%$. A redução física das horas formais é a maior parcela negativa da decomposição; a resposta da eficiência depende da especificação, enquanto a realocação para o emprego informal amplia a perda.

Os contrafactuais também separam três objetos que não devem ser confundidos: a perda de produto com PTF fixa, o ganho de PTF que recompõe essa perda e o indicador GHH de consumo e lazer. Reduções moderadas podem diminuir o produto e, simultaneamente, elevar o indicador representativo de bem-estar. No teto de 36 horas, o requisito de PTF é quantitativamente maior e o equivalente de consumo é negativo nas duas funções de eficiência, variando de $-3{,}42\%$ a $-1{,}66\%$. As sensibilidades mostram que a avaliação depende da fadiga e da distribuição inicial de horas. A base de 44h é uma hipótese de mensuração da reforma, distinta de horas contratadas observadas.

As conclusões são condicionais a capital e emprego agregados predeterminados. Incorporar investimento, barganha, seleção de trabalhadores, ligações setoriais e respostas de oferta de trabalho é a extensão natural para uma avaliação completa. Dentro do objeto estudado, os resultados mostram que a intensidade da redução e a resposta da produtividade alteram de forma substantiva sua aritmética macroeconômica e motivam a avaliação de trajetórias graduais e de instrumentos de transição.

\FloatBarrier

\bibliographystyle{chicago}
\bibliography{references}

\end{document}